\documentclass{llncs}
\usepackage[T1]{fontenc}
\usepackage[hidelinks]{hyperref}
\usepackage{amsmath}
\usepackage{float}
\usepackage{multirow}
\usepackage{graphicx}
\usepackage[normalem]{ulem}
\usepackage[linesnumbered,ruled,vlined]{algorithm2e}
\usepackage{booktabs}
\usepackage{color}
\usepackage{comment}
\usepackage{enumitem}
\usepackage{multirow}
\usepackage{subcaption}

\newcommand{\etal}{\textit{et al.}}
\newcommand{\uigccf}{\textit{UI-GCCF}\xspace}

\begin{document}
\titlerunning{Item Graph Convolution Collaborative Filtering}

\title{Item Graph Convolution Collaborative Filtering for Inductive Recommendations}

\author{Edoardo D'Amico\orcidID{0000-0002-8262-7207} \and Khalil Muhammad\orcidID{0000-0002-8755-7535} \and Elias Tragos\orcidID{0000-0001-9566-531X} \and Barry Smyth\orcidID{0000-0003-0962-3362} \and Neil Hurley\orcidID{0000-0001-8428-2866} \and Aonghus Lawlor\orcidID{0000-0002-6160-4639}}

\authorrunning{E. D'Amico et al.}
\institute{Insight Centre for Data Analytics, Dublin, Ireland\\
\email{\{name.surname\}@insight-centre.org}}
\maketitle  
\begin{abstract}
Graph Convolutional Networks (GCN) have been recently employed as core component in the construction of recommender system algorithms, interpreting user-item interactions as the edges of a bipartite graph.
However, in the absence of \textit{side information}, the majority of existing models adopt an approach of randomly initialising the user embeddings and optimising them throughout the training process.
This strategy makes these algorithms inherently \textit{transductive}, curtailing their ability to generate predictions for users that were unseen at training time. 
To address this issue, we propose a convolution-based algorithm, which is \textit{inductive} from the user perspective, while at the same time, depending only on implicit user-item interaction data.
We propose the construction of an item-item graph through a weighted projection of the bipartite interaction network and to employ convolution to inject higher order associations into item embeddings, while constructing user representations as weighted sums of the items with which they have interacted.
Despite not training individual embeddings for each user our approach achieves state-of-the-art recommendation performance with respect to \textit{transductive} baselines on four real-world datasets, showing at the same time robust inductive performance. 
\end{abstract}


\section{Introduction}
Recent years have witnessed the success of Graph Convolutional Networks based algorithm in many domains, such as social networks \cite{kipf2016semisupervised,chen2018fastgcn}, natural language processing \cite{yao2019graph} and computer vision \cite{wang2018zero}. The core component of Graph Convolutional Networks algorithms is the iterative process of aggregating information mined from node
neighborhoods, with the intent of capturing high-order associations between nodes in a graph. 
GCNs have opened a new perspective for recommender systems in light of the fact that user-item interactions can be interpreted as the edges of a bipartite graph \cite{wang2019neural,Chen_2020,He_2020}. 
Real-world recommender system scenarios must contend with the issue that user-item graphs change dynamically over time. New users join the system on a daily basis, and existing users can produce additional knowledge by engaging with new products (introducing new edges in the user-item interaction graph). The capacity to accommodate new users to the system — those who were not present during training — and fast leverage novel user-item interactions is a highly desirable characteristic for recommender systems meant to used in real-world context. Delivering high quality recommendations 
under these circumstances poses a severe problem for many existing \textit{transductive} recommender system algorithms. Models such as \cite{wang2019neural,Chen_2020,He_2020} need to be completely re-trained to produce the embedding for a new user that joins the system post-training and the same happens when new user-item interactions must be considered; this limitation restricts their use in real-world circumstances. \cite{localfactormodels}.

One solution present in literature, is to leverage side information (user and item metadata) beyond the pure user-item interactions in order to learn a mapping function from user and item features to embeddings \cite{volkovs2017dropoutnet,zhang2019inductive,jain2013provable,pmlr-v80-hartford18a}.  However, it can be difficult to obtain this additional side information in many real-world scenarios, as it may be hard to extract, unreliable, or simply unavailable. For example, when new users join a system, there may be very little or no information available about them, making it difficult or impossible to generate their embeddings. Even when it is possible to gather some information about these users, it may not be useful in inferring their preferences. Another way to account for new users and rapidly create embeddings which exploit new user-item interactions is to resort to \textit{item-based} models \cite{Cremonesi2010Performance,kabbur2013fism}. In this setting only the item representations are learnt and then exploited to build the user embeddings. Anyway these category of models do not directly exploit the extra source of information present in the user-item interaction graph, which have been shown to benefit the performance of the final model. Furthermore the application of standard Graph Convolution methods recently presented for the collaborative filtering problem have not been extended to work in a setting where only the item representations are learnt.

In this paper we propose a novel item-based model named Item Graph Convolutional Collaborative Filtering (IGCCF), capable of handling dynamic graphs while also leveraging the information contained in the user-item graph through graph convolution. It is designed to learn rich item embeddings capturing the higher-order relationships existing among them. To extract information from the user-item graph we propose the construction of an item-item graph through a weighted projection of the bipartite network associated to the user-item interactions with the intent of mining high-order associations between items. We then construct the user representations as a weighted combination of the item embeddings with which they have previously interacted, in this way we remove the necessity for the model to learn static one-hot embeddings for users, reducing the space complexity of previously introduced GCN-based models and, at the same time, unlocking the ability to handle dynamic graphs, making straightforward the creation of the embeddings for new users that join the system post training as well as the ability of updating them when new user-item interactions have been gathered, all of that without the need of an expensive retraining procedure.

\section{Preliminaries and Related Work}
In this paper we consider the extreme setting for inductive recommendation in which user preferences are estimated by leveraging only past user-item interactions without any additional source of information. We focus on implicit user feedback \cite{rendle2012bpr}, with the understanding that explicit interactions are becoming increasingly scarce in real-world contexts. More formally, denoting with $\mathcal{U}$ and $\mathcal{I}$ the sets of users and items, and with $U = |\mathcal{U}|$ and $I = |\mathcal{I}|$ their respective cardinalities, we define the user-item interaction matrix $\textbf{R}_{U \times I}$, where cell $r_{ui} = 1$ if user $u$ has interacted with item $i$, and $0$ otherwise, as the only source of information.



\subsection{GCN-based Recommender}
GCN-based models have recently been applied to recommender system models, by virtue of the fact that historical user-item interactions can be interpreted as the edges of a graph.  It is possible to define the adjacency matrix $\textbf{A}$, associated with an undirected bipartite graph, exploiting the user-item interaction matrix $\textbf{R}_{U \times I}$, as:
\begin{align*}
    \textbf{A} = 
    \begin{bmatrix}
    \textbf{0}_{U \times I} & \textbf{R} \\
    \textbf{R}^T & \textbf{0}_{I \times U}
    \end{bmatrix}
\end{align*}
The set of the graph's nodes is $\mathcal{V} = \mathcal{U} \bigcup \mathcal{I}$ and there exists an edge between a user $u$ and an item $i$ if the corresponding cell of the interaction matrix $r_{ui} = 1$.
He et al. \cite{wang2019neural}, first applied graph convolution in a setting where no side information was available, and proposed to initialise the node representations with free parameters. This formulation is a variant of the one proposed in \cite{kipf2016semisupervised} but includes information about the affinity of two nodes, computed as the dot product between embeddings. 
Subsequently, Chen et al.~\cite{Chen_2020} have shown how the affinity information as well as the non-linearities tend to complicate the training process as well as degrade the overall performance.
Finally, He et al. \cite{He_2020}, confirmed the results of \cite{wu2019simplifying} by showing how the benefits of graph convolution derive from smoothing the embeddings and that better performance can be achieved by removing all the intermediary weight matrices. In this formulation, the embeddings of users and items at depth $k$ can be simply computed as the linear combination of the embeddings of the previous step with weights assigned from a suitably chosen propagation matrix \textbf{P}.

\subsection{Item-based Recommender}
Item-based models aim to learn item embeddings which are subsequently used to infer user representations. As a result, this model category is capable of providing recommendations to new users who join the system post training.
Cremonesi \etal \cite{Cremonesi2010Performance}, proposed PureSVD which uses singular value decomposition to retrieve item representations from the user-item interaction matrix, and subsequently compute the user embeddings as a weighted combination of item representations.
Later, Kabbur \etal in \cite{kabbur2013fism} also propose 
to compute users as a weighted combination of items, but instead of 
computing them after the creation of the item embeddings, they are jointly used together with the item representation as part of an optimisation process. 

Our proposed IGCCF model inherits from the item-based model the core idea of inferring user embeddings from items, but it is also capable of leveraging the information contained in the graph-structure during the item representation learning phase through graph convolution.

\section{Methodology}
\begin{figure*}[!t]
    \centering
    \includegraphics[angle=0,width=1\textwidth]{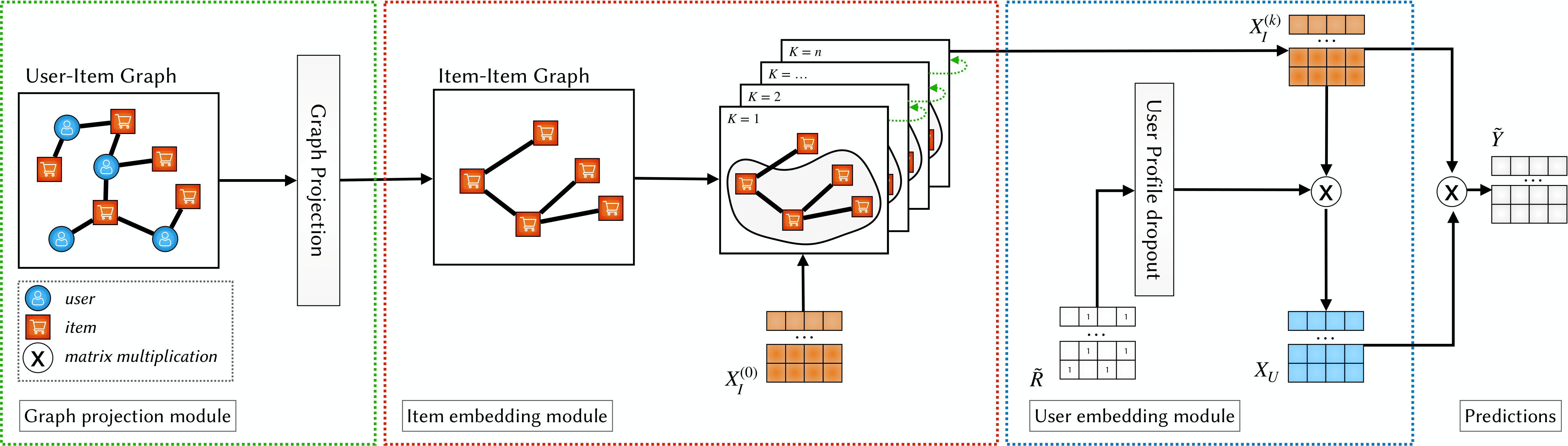}
\caption{Model architecture.}
\label{fig:igcmf_architecture}
\end{figure*}
In this section we present details of the proposed model. \textit{IGCCF} comprises three different elements: (1) a graph projection module, which is used to transform a user-item bipartite graph into a homogeneous item-item graph; (2) an item embedding module, which is used to learn item embeddings starting from the item-item graph; (3) a user embedding module, which is used to build user embeddings given the user-item interaction matrix and the items embeddings. 
The overall architecture is presented in Figure \ref{fig:igcmf_architecture}. 

\subsection{Graph projection Module}
The graph convolution module operates over item embeddings which are optimised during training while the explicit representation and optimisation of separate user embeddings is not required. This gives the model the flexibility to easily make recommendations for unseen users. To fully capture the item relationships 
we construct an item-item relational graph from which extract knowledge regarding item associations during the representation learning process.
The purpose of the graph projection module is to transform the bipartite user-item graph into a homogeneous item-item graph. The simplest means of achieving this is to use a one-mode projection onto the set of item nodes $\mathcal{I}$, creating an unweighted graph with exactly $I$ nodes where two item nodes share an edge when they have at least one common neighbour in $\mathcal{U}$ \cite{Newman404}. 
This technique ignores the frequency with which two nodes share neighbors, resulting in information loss. To account for this, we build the projected item-item graph by weighting the edges based on the cosine similarity of item profiles. The edge between nodes $i$ and $j$ has weight $w_{ij} =\frac{ \textbf{r}_i \cdot \textbf{r}_j} {||\textbf{r}_i|| \cdot ||\textbf{r}_j||}$ where $i, j \in \mathcal{I}$ and indicating with $\textbf{r}_i$ the $i^{th}$ column of the matrix $\textbf{R}$. In this way we are able to retain information about the frequency with which two items share neighbors.
The model can easily adapt to different bipartite graph projection methodologies such as hyperbolic weighting \cite{newman2001scientific} that takes into account the saturation effect; or weighting based on resource allocation \cite{zhou2007bipartite}, which doesn't assume symmetric weights between pairs of nodes.  

\subsubsection{Top-K pruning}
Previous works on GCNs have highlighted how the size of the neighbourhood included in the convolution operation, as well as the convolution depth, can lead to an \textit{oversmoothing} of the embeddings.
The oversmoothing leads to a loss of embedding uniqueness, and results in the degradation of recommendation performance \cite{li2018deeper,xu2018representation,Chen_2020}. To address this problem we apply a top-K pruning preprocessing step on the edges of the item-item graph, keeping only the $K$ edges associated to the highest similarity score, for each item node. In this way only the most important neighbours are included in every convolution operation reducing the effect of the smoothing phenomenon. In section~\ref{sec:exp_topk_pruning} we show how the top-K pruning is beneficial to both training time and recommendation performance of the presented algorithm.

\subsection{Item embedding module}
The item embedding module uses information from the item-item graph to generate refined item embeddings. The primary difference between this module and previously described graph convolution modules  \cite{wang2019neural,Chen_2020,He_2020} is that we use the item-item similarity matrix as propagation matrix, allowing us to directly leverage the information provided by the weighted projection used to construct the homogeneous item graph.


  
At the first iteration, $k=0$, the item embedding matrix $\textbf{X}^{(0)}$ is randomly initialised. 
%
At each subsequent iteration $k$, the item embedding matrix is a weighted combination of the embedding matrix at the previous layer $k-1$ with the propagation matrix, formed from the cosine similarity measure:
\begin{equation}
\textbf{X}^{(k)} = \textbf{P} \textbf{X}^{(k-1)} = \textbf{P}(\textbf{P}\textbf{X}^{(k-2)}) = \textbf{P}^k\textbf{X}^0\
\label{eq:power_prop_matrix}
\end{equation}
The representation of an item $i$ at convolution depth $k$ can be written explicitly as:
\begin{align*}
\textbf{x}_{i}^{(k)} = \sum_{j \in \mathcal{N}_i} w_{ij}\textbf{x}_j^{(k-1)}
\end{align*}
where $\mathcal{N}_i$ represents  the 1-hop neighbourhood of item $i$. 

The embedding at depth $k$ can  be directly computed using the power of the propagation matrix as shown in \autoref{eq:power_prop_matrix}, which demonstrates that, at depth $k$, the embedding can be seen as the linear combination of neighbourhoods representations up to $k$-hop distance with weights given by the $k^{th}$ power of the cosine similarity matrix $\textbf{P}^k$.

\subsection{User embedding module}
As there are no separate user embeddings, a method to map users into the item embedding space is required. We propose to map a user inside the item latent space as a weighted combination of the items in their profile. Given the item embeddings,  a user embedding is created as:
\begin{equation}
\label{eqn:userembedding}
\textbf{x}_u = \sum_{i \in \mathcal{I}} \lambda_{ui} r_{ui}  \textbf{x}_i
\end{equation}
where 
$\lambda_{ui}$ is a scalar weighting the contribution of item $i$ to the embedding of user $u$ and $\textbf{x}_i $ represents the embedding of item $i$. 
We can compute the user embeddings in matrix form as follows:
\begin{align*}
\textbf{U} = (\textbf{R} \odot \Lambda) \textbf{X} = \tilde{\textbf{R}}\textbf{X}   
\end{align*}
where $\odot$ indicates the Hadamard product, $\tilde{\textbf{R}}$ represents a weighted version of the interaction matrix and $\textbf{X}$ is the item embedding matrix. In the proposed work, we assign uniform weights to all user interactions and leave the investigation of different weighting mechanisms as future work. 

We want to emphasize the key advantages of modeling a user as a weighted sum of item embeddings in their profiles over having a static one-hot representation for each of them.
First, it makes the model inductive from the user perspective and endows IGCCF with the ability to perform real-time updates of the user-profile as it is possible to create the embedding of a new user as soon as they start interacting with items in the system using \autoref{eqn:userembedding}.
Second, it improves the model's space complexity from $\mathcal{O}(I+U)$ to $\mathcal{O}(I)$ when compared to transductive models. Finally, different importance scores may be assigned to user-item interactions when generating the user embeddings, this might be beneficial in situations where recent interactions are more significant than older ones.

\subsection{Model Training}
To learn the model parameters, we adopt the \textit{Bayesian Personalised Ranking} (BPR) loss \cite{rendle2012bpr}:
\begin{align*}
L_{BPR} = \sum_{(u,i^+,i^-)\in \mathcal{O}} -\ln \sigma (\hat{y}_{ui^+}-\hat{y}_{ui^-}) + \lambda ||\Theta||_2^2
\end{align*}
where $\mathcal{O}=\{(u, i^+, i^-)|(u, i^+) \in \mathcal{R^+}, (u, i^-) \in \mathcal{R^-}\}$ denotes the pairwise training data, $\mathcal{R^+}$ indicates the observed interactions, and $\mathcal{R^-}$ the unobserved interactions; $\sigma(\cdot)$ represents the sigmoid activation function; $\Theta$ are the parameters of the model which correspond to the item embeddings.

We use the Glorot initialisation for the item embeddings \cite{glorot2010understanding} and mini-batch stochastic gradient descent with Adam as optimiser \cite{kingma2014adam}. The preference of a user for an item is modelled through the standard dot product of their embeddings $\hat{y}(u, i) = \textbf{x}_u^T \cdot \textbf{x}_i$

\subsubsection{User-profile dropout}
It is well-known that machine learning models can suffer from overfitting. Following previously presented works on GCNs \cite{rong2019dropedge,wang2019neural,berg2017graph}, 
we design a new dropout mechanism called \textit{user-profile} dropout. Before applying ~\autoref{eqn:userembedding} to form the user embeddings,  we randomly drop entries of the weighted user interaction matrix $\tilde{R}$ with probability $p \in [0,1]$. The proposed regularisation mechanism is designed to encourage the model to rely on strong patterns that exist across items rather than allowing it to focus on a single item during the construction of user embeddings.
\section{Experiments}
We perform experiments on four real-world datasets to evaluate the proposed model. We answer to the following research questions. \textbf{[RQ1]}: How does IGCCF perform against transductive graph convolutional algorithms?
\textbf{[RQ2]}: How well does IGCCF generalise to unseen users?
\textbf{[RQ3]}: How do the hyperparameters of the algorithm affect its performance?

\subsection{Datasets}
To evaluate the performance of the proposed methodology we perform experiments on four real world datasets gathered in different domains. \textbf{LastFM}: Implicit interactions from the Last.fm music website. In particular, the user \textit{listened} artist relation expressed as listening counts \cite{cantador2011second}. We consider a positive interaction as one where the user has listened to an artist. \textbf{Movielens1M}: User ratings of movies from the MovieLens website \cite{movielens}. Rating values range from $1$ to $5$, we consider ratings $\geq$ 3 as positive interactions. \textbf{Amazon Electronics}: User ratings of electronic products from the Amazon platform \cite{He_2016,mcauley2015image}. The rating values also range from $1$ to $5$, so we consider ratings $\geq$ 3 as positive interactions. \textbf{Gowalla} User \textit{check-ins} in key locations from Gowalla \cite{liang2016modeling}.
Here, we consider a positive interaction between a user and a location, if the user has checked-in at least once. To ensure the integrity of the datasets, following \cite{He_2020,wang2019neural}, we perform a \textit{k}-core preprocessing step setting $k_{core} = 10$, meaning we discard all users and items with less than ten interactions.
\begin{table*}[!t]
\small
\setlength{\tabcolsep}{1pt}
\caption{Transductive performance comparison. Bold and underline indicate the first and second best performing algorithm respectively.}
\centering
\begin{tabular}{ccccc@{\hspace{10pt}}cccc@{\hspace{10pt}}}
\toprule 
 &
\multicolumn{4}{c}{\textbf{LastFM}} & 
\multicolumn{4}{c}{\textbf{Ml1M}} \\
user/item/int&\multicolumn{4}{c}{1,797/1,507/6,376}&\multicolumn{4}{c}{6,033/3,123/834,449}\\

\cmidrule(l{15pt}r{20pt}){2-5}
\cmidrule(l{15pt}r{20pt}){6-9}
\multirow{2}{*}{Model}& 
 \multicolumn{2}{c}{NDCG} & 
 \multicolumn{2}{c}{Recall} &
 
 \multicolumn{2}{c}{NDCG} & 
 \multicolumn{2}{c}{Recall} \\
 
 

 & 
 @5 & @20 & @5 & @20 &
 @5 & @20 & @5 & @20 \\ \midrule

 BPR-MF &
 0.2162 & 0.3027 & 0.2133 & 0.4206 &
 0.1883 & 0.3173 & 0.1136 & 0.2723  \\ 

 iALS &
 0.2232 & 0.3085 & 0.2173 & 0.4227 &
 \underline{0.2057} & \underline{0.3410} & \underline{0.1253} & \underline{0.2893}  \\ \midrule

  PureSVD&
 0.1754 & 0.2498 & 0.1685 & 0.3438 &
 0.2024 & 0.3369 & 0.1243 & 0.2883 \\

   FISM&
 0.2143 & 0.2978 & 0.2145 & 0.4139 &
 0.1929 & 0.3188 & 0.1203 & 0.2805 \\ \midrule
 
 NGCF &
 0.2216 & 0.3085 & 0.2185 & 0.4299 &
 0.1996 & 0.3309 & 0.1206 & 0.2821  \\

 LightGCN &
 \underline{0.2293} & \underline{0.3157} & \underline{0.2287} & \underline{0.4379} &
 0.1993 & 0.3319 & 0.1218 & 0.2864  \\ \midrule

 IGCCF (Ours)&
 \textbf{0.2363} & \textbf{0.3207} & \textbf{0.2372} & \textbf{0.4405} &
 \textbf{0.2070} & \textbf{0.3456} & \textbf{0.1249} & \textbf{0.2954} \\
\midrule
\midrule


 &

\multicolumn{4}{c}{\textbf{Amazon}} &
\multicolumn{4}{c}{\textbf{Gowalla}} \\
user/item/int&\multicolumn{4}{c}{13,455/8,360/234,521}&\multicolumn{4}{c}{29,858/40,988/1,027,464}\\

\cmidrule(l{15pt}r{20pt}){2-5}
\cmidrule(l{15pt}r{20pt}){6-9}
\multirow{2}{*}{Model}& 
 \multicolumn{2}{c}{NDCG} & 
 \multicolumn{2}{c}{Recall} &
 
 \multicolumn{2}{c}{NDCG} & 
 \multicolumn{2}{c}{Recall} \\
 
 & 
 @5 & @20 & @5 & @20 &
 @5 & @20 & @5 & @20 \\ \midrule
 
 BPR-MF &
 0.0247 & 0.0419 & 0.0336 & 0.0888  &
 0.0751 & 0.1125 & 0.0838 & 0.1833 \\ 

iALS &
 0.0273 & 0.0432 & 0.0373 & 0.0876 &
 0.0672 & 0.1013 & 0.0763 & 0.1667  \\ \midrule

  PureSVD&
 0.0172 & 0.0294 & 0.0244 & 0.0631 &
 0.0795 & 0.1032 & 0.0875 & 0.1861 \\

   FISM&
 0.0264 & 0.0424 & 0.0353 & 0.0865 &
 0.0812 & 0.1191 & 0.0915 & 0.1925 
 \\ \midrule
 
 NGCF &
 0.0256 & 0.0436 & 0.0346 & 0.0926  &
 0.0771 & 0.1156 & 0.0867 & 0.1896 \\

 LightGCN &
 \underline{0.0263} & \underline{0.0455} & \underline{0.0358} & \underline{0.0978}  &
 \underline{0.0874} & \underline{0.1279} & \underline{0.0975} & \underline{0.2049}
 \\ \midrule

 IGCCF (Ours)&
\textbf{0.0336} & \textbf{0.0527} & \textbf{0.0459} & \textbf{0.1072} &
 \textbf{0.0938} & \textbf{0.1373} & \textbf{0.1049} & \textbf{0.2203}
 \\ 
 
 \bottomrule

\end{tabular}
\label{table:performance_comparison}
\end{table*}
\subsection{Baselines}
To demonstrate the benefit of our approach we compare it against the following baselines: \textbf{BPRMF} \cite{rendle2012bpr} Matrix factorisation optimised by the BPR loss function. \textbf{iALS} \cite{hu2008collaborative} matrix factorization learned by implicit alternating least squares. \textbf{PureSVD}\cite{Cremonesi2010Performance}Compute item embeddings through a singular value decomposition of the user-item interaction matrix, which will be then used to infer user representations.  \textbf{FISM}\cite{kabbur2013fism} Learn item embeddings through optimisation process creating user representations as a weighted combination of items in their profile. Additional user and item biases as well as an agreement term are considered in the score estimation. \textbf{NGCF} \cite{wang2019neural} Work that introduces graph convolution to the collaborative filtering scenario, it uses dense layer and inner product to enrich the knowledge injected in the user item embeddings during the convolution process.
\textbf{LightGCN} \cite{He_2020} Simplified version of graph convolution applied to collaborative filtering directly smooth user and item embeddings onto the user-item bipartite graph. We follow the original paper \cite{He_2020} and use $a_k = 1/(k+1)$.

For each baseline, an exhaustive grid-search has been carried out to ensure optimal performance. Following \cite{He_2020}, for all adopted algorithms the batch size has been  set to $1024$ and embedding size to $64$.
Further details on the ranges of the hyperparameter search as well as the data used for the experiments are available in the code repository \footnote{\href{https://github.com/damicoedoardo/IGCCF}{https://github.com/damicoedoardo/IGCCF}}.

\subsection{Transductive performance}
\label{sec:transductive_perf}
In this section we evaluate the performance of IGCCF against the proposed baselines in a transductive setting, meaning considering only users present at training time.
To evaluate every model, following \cite{He_2020,wang2019neural}, for each user, we randomly sample  $80\%$ of his interactions to constitute the training set, $10\%$ to be the test set, while the remaining $10\%$ are used as a validation set to tune the algorithm hyper-parameters. Subsequently, validation and training data are merged together and used to retrain the model, which is then evaluated on the test set.
In order to asses the quality of the recommendations produced by our system, we follow the approach outlined in
\cite{wang2019neural,wu2019simplifying,Chen_2020}. For each user in the test data, we generate a ranking of items and calculate the average \textit{Recall@N} and \textit{NDCG@N} scores across all users, considering two different cutoff values $N = 5$ and $N = 20$. The final results of this analysis are presented in \autoref{table:performance_comparison}. 

Based on the results obtained, we can establish that IGCCF outperforms NGCF and LightGCN on all four datasets examined for each metric and cutoff. This confirms that explicitly parametrizing the user embeddings is not necessary to get the optimum performance; on the contrary, it might result in an increase in the number of parameters of the model, which is detrimental to both training time and spatial complexity of the model.
Furthermore, IGCCF shows superior performance with respect to the item-based baseline models. This demonstrates that interpreting user-item interaction as graph-structured data introduces relevant knowledge into the algorithm learning process, leading to improved model performance.
\subsection{Inductive performance}

\begin{table*}[!t]
\small
\caption{Inductive performance on \emph{unseen} users. Bold indicates the performance of the best ranking algorithm.}
\centering
\begin{tabular}{ccccc@{\hspace{10pt}}cccc@{\hspace{10pt}}}
\toprule 

\multirow{3}{*}{Model} &

\multicolumn{4}{c}{LastFM} & 
\multicolumn{4}{c}{Ml1M} \\


\cmidrule(l{15pt}r{20pt}){2-5}
\cmidrule(l{15pt}r{20pt}){6-9}
& 
 \multicolumn{2}{c}{NDCG} & 
 \multicolumn{2}{c}{Recall} &
 
 \multicolumn{2}{c}{NDCG} & 
 \multicolumn{2}{c}{Recall} \\
 
 

 & 
 @5 & @20 & @5 & @20 &
 @5 & @20 & @5 & @20 \\ \midrule
 
  PureSVD&
 0.1640 & 0.2279 & 0.1610 & 0.3124 &
 0.2064&0.3418&0.1165&0.2759 \\

   FISM&
 0.1993 & 0.2921 & 0.1927 & 0.4165 &
 0.1974&0.3221&0.1105&0.2638 \\
 
 IGCCF (Ours)&
 \textbf{0.2374} & \textbf{0.3227} & \textbf{0.2355} & \textbf{0.4395} &
 \textbf{0.2089} & \textbf{0.3474} & \textbf{0.1177} & \textbf{0.2817} \\
\midrule
\midrule


\multirow{3}{*}{Model} &

\multicolumn{4}{c}{Amazon} &
\multicolumn{4}{c}{Gowalla} \\

\cmidrule(l{15pt}r{20pt}){2-5}
\cmidrule(l{15pt}r{20pt}){6-9}
& 
 \multicolumn{2}{c}{NDCG} & 
 \multicolumn{2}{c}{Recall} &
 
 \multicolumn{2}{c}{NDCG} & 
 \multicolumn{2}{c}{Recall} \\
 
 & 
 @5 & @20 & @5 & @20 &
 @5 & @20 & @5 & @20 \\ \midrule

  PureSVD&
 0.0221 & 0.0345 & 0.0320 & 0.0721 &
 0.0815 & 0.1213 & 0.0862 & 0.1910 \\
  
   FISM&
 0.0330 & 0.0468 & 0.0424 & 0.0891 &
 0.0754 & 0.1102 & 0.0829 & 0.1763 
 \\ 
 
 IGCCF (Ours)&
\textbf{0.0356} & \textbf{0.0513} & \textbf{0.0477} & \textbf{0.0978} &
 \textbf{0.0910} & \textbf{0.1341} & \textbf{0.1009} & \textbf{0.2172}
 \\ 
 
 \bottomrule

\end{tabular}
\label{table:inductive_performance_comparison}
\end{table*}
\label{sec:inductiveperf}
A key feature of the proposed IGCCF algorithm, is the ability to create embeddings and consequently retrieve recommendations for \emph{unseen} users who are not present at training time. IGCCF does not require an additional learning phase to create the embeddings. As soon as a new user begins interacting with the items in the catalogue, we may construct its embedding employing \autoref{eqn:userembedding}.

To assess the inductive performance of the algorithm we hold out $10\%$ of the users, using the remaining $90\%$ as training data.
For every unseen user we use $90\%$ of their profile interactions to create their embedding (Eq. \ref{eqn:userembedding}) and we evaluate the performance on the remaining $10\%$ of interactions.
We compare the performance of our model against the inductive baselines corresponding to the item-based models (PureSVD and FISM) since the transductive models are not able to make predictions for users who are not present at training time without an additional learning phase.
Recommendation performance is evaluated using the same metrics and cutoffs reported in \autoref{sec:transductive_perf}. The overall results are reported in \autoref{table:inductive_performance_comparison}.
IGCCF outperforms the item-based baselines on all the datasets. These results strongly confirm our insight that the knowledge extracted from the constructed item-item graph is beneficial to the item-embedding learning phase, even when making predictions for unseen users.

\subsubsection{Robustness of inductive performance}
 \begin{figure}
    \centering
    \includegraphics[width=1\textwidth, keepaspectratio]{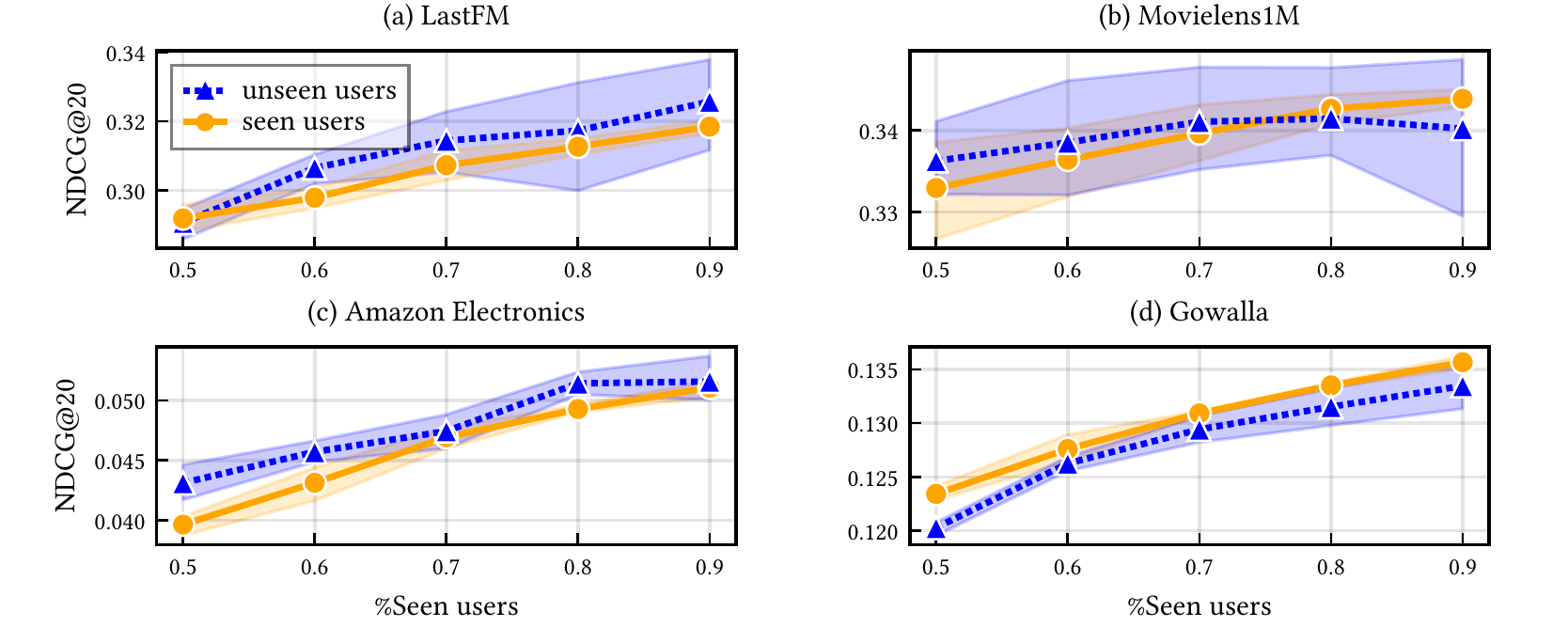}
    
    \caption{For each dataset we vary the percentage of users in the training data, and evaluate the performance of IGCCF on both \textit{seen} and \textit{unseen} users.}
    
    \label{fig:Inductive_performance}    
\end{figure}

We are interested in the extent to which IGCCF can maintain comparable recommendation performance between \emph{seen} and \emph{unseen} users as we train the model with less data. For this experiment, we increasingly reduce the percentage of seen users which are used to train the model and consequently increase the percentage of unseen users which are presented for inductive inference. We train the model 5 times on each different split used and we report the average performance (NDCG@20). From the results in \autoref{fig:Inductive_performance} we can observe:
IGCCF exhibits comparable performance on both seen and unseen user groups for all the splits analysed, showing how the inductive performance of IGCCF is robust with respect to the amount of training data available. As expected, reducing the amount of available training data results in a lower $NDCG@20$, anyway is interesting to notice how the drop in performance is minimal even when the model is trained with half of the data available.

\begin{figure}
    \centering
    \includegraphics[width=\textwidth, keepaspectratio]{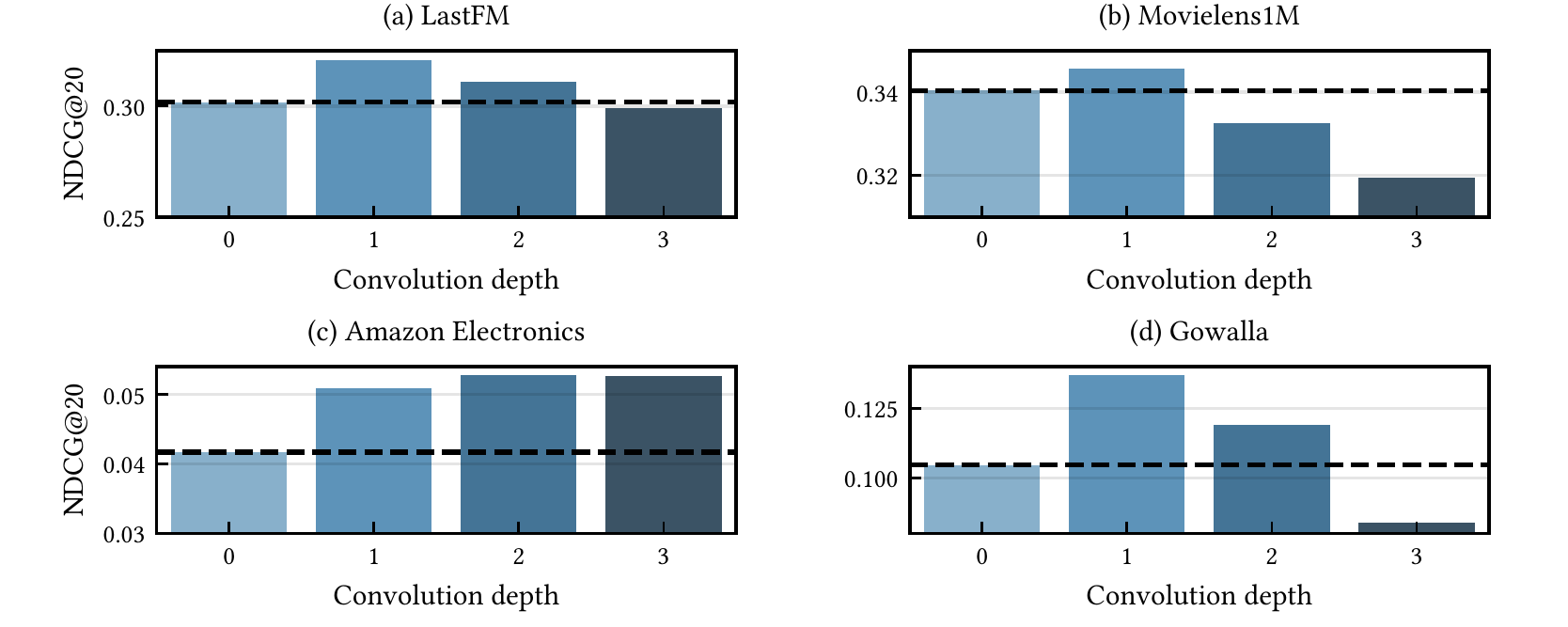}
    
    \caption{Ablation study: Effect of the Convolution depth parameter on the algorithm performance.}
    \label{fig:ablation_cdepth}
\end{figure}
\subsection{Ablation study}
\subsubsection{Convolution Depth}
The convolution operation applied during the learning phase of the item embeddings, is beneficial in all the studied datasets, the results are reported in  \autoref{fig:ablation_cdepth}. It is interesting to consider the relationship between the dataset density and the effect of the convolution operation. We can see that the largest improvement of $31\%$ is found on Gowalla, which is the least dense dataset ($0.08\%$). As the density increases, the benefit introduced by the convolution operation decreases. We have an improvement of $26\%$ and $6\%$ on Amazon Electronics ($0.21\%$) and LastFM ($2.30\%$) respectively
while there is a very small increase of $1.5\%$ on Movielens1M ($4.43\%$). 
The results obtained suggest an inverse correlation between the dataset density and the benefit introduced by the convolution operation.
\subsubsection{User-profile dropout}
\begin{figure}
    \centering
    \includegraphics[width=\textwidth, keepaspectratio]{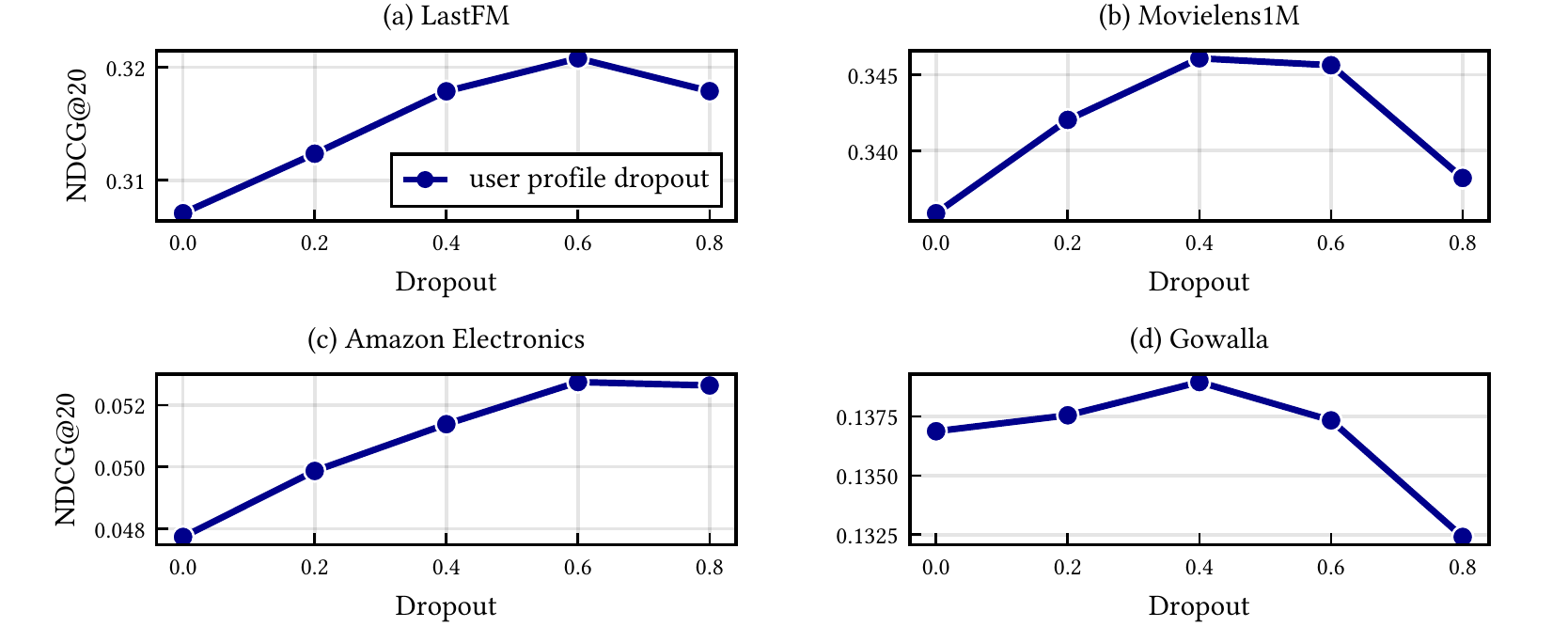}
    
    \caption{Ablation study: Effect of the dropout mechanisms on the algorithm performance.}
    \label{fig:ablation_dropout}
\end{figure}
From the analysis reported in \autoref{fig:ablation_dropout}, it is clearly visible that user profile dropout regularisation have a strong impact on the performance of the proposed method. In all four datasets, the utilisation of the suggested regularisation technique 
enhance the quality of the recommendation performance, resulting in a gain over the \textit{NDCG@20} metric of 4.4\%, 3.0\%, 10.5\%, 1.5\% for LastFM, Movielens1M, Amazon Electronics and Gowalla respectively. Dropping a portion of the user profiles during the embeddings creation phase, force the algorithm to not heavily rely on information coming from specific items.

\subsubsection{Top-K pruning}
\begin{figure}
    \centering
    \includegraphics[width=\textwidth, keepaspectratio]{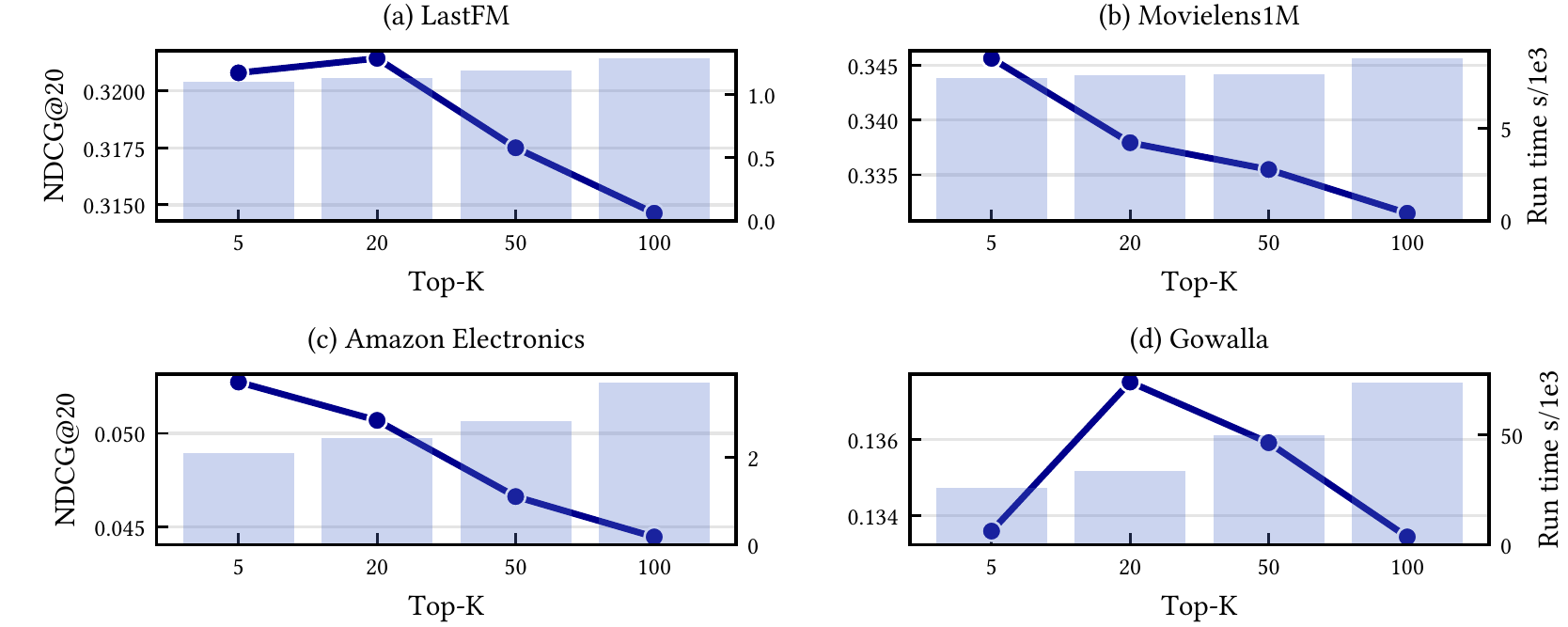}
    
    \caption{Effect of the top-K preprocessing step on the algorithm performance and training time.}
    \label{fig:ablation_top-k}
\end{figure}
\label{sec:exp_topk_pruning}
To prevent the well-known oversmoothing issue caused by graph convolution, we trim the edges of the item-item graph to maintain only the most strong connections between items. \autoref{fig:ablation_top-k} illustrates the results of the ablation study. In all of the datasets investigated, utilising at most $20$ neighbours for each item node yields the highest performance; this demonstrates how retaining edges associated with weak item links can worsen model performance while also increasing the algorithm training time.

\section{Conclusion and Future work}
In this work we presented IGCCF, an item-based model that employs graph convolution to learn refined item embeddings. We build upon the previously presented graph convolution models by removing the explicit parameterisation of users. The benefits of that are threefold: first, it reduces model complexity; second, it allows real-time user embeddings updates as soon as new interactions are gathered; and third, it enables inductive recommendations for new users who join the system post-training without the need for a new expensive training procedure. To do this, we devised a novel procedure that first constructs an item-item graph from the user-item bipartite network. A top-K pruning procedure is then employed to refine it, retaining only the most informative edges. Finally, during the representation learning phase, we mine item associations using graph convolution, building user embeddings as a weighted combination of items with which they have interacted. In the future, we will extend the provided methodology to operate in settings where item side-information are available.
\bibliographystyle{splncs04}
\bibliography{bibliography.bib}

\end{document}